\documentclass[conference,a4paper]{APSIPA2021}
\usepackage{multirow}
\usepackage[dvips]{graphicx}
\usepackage{amsmath}
\usepackage[psamsfonts]{amssymb}
\usepackage{amsxtra}
\usepackage{threeparttable}

\usepackage[export]{adjustbox}
\usepackage{float}

\begin{document}

\title{Design and Evaluation of Active Noise Control on Machinery Noise}

\author{%
\authorblockN{%
Shulin Wen\authorrefmark{1},
Duy Hai Nguyen,
Miqing Wang, and
Woon-Seng Gan
}
\authorblockA{%
\authorrefmark{1}
Digital Signal Processing Lab, School of Electrical and Electronic Engineering 
Nanyang Technological University, Singapore \\
E-mail: alicia.wen@ntu.edu.sg,\\
NDHai@ntu.edu.sg,\\
miqing001@e.ntu.edu.sg,\\
EWSGAN@ntu.edu.sg}
}

\maketitle
\thispagestyle{empty}

\begin{abstract}
  Construction workers and residents live near around construction sites are exposed to noises that might cause hearing loss, high blood pressure, heart disease, sleep disturbance and stress. Regulations has been carried out by national governments to limit the maximum permissible noise levels for construction works. A four-channel active noise control system mounted on the opening of an enclosure is designed to prevent the machinery noise from spreading around and retaining the heat diffusion path. Multi-channel FxLMS algorithm in time domain is implemented on the main controller. A Genelec speaker is placed inside the box as the primary noise source to play back different types of noises. Analyses and experiments are carried out to investigate the controllable frequency range of this ANC system in detail. Considerable noise reduction performance is achieved for different recorded practical construction noises.
\end{abstract}

\section{Introduction}
With the growth of urbanization, noise pollution has become one of the major threats to the quality of our lives in modern society. Construction works are inevitable in urban cities, which produces massive high sound pressure level (SPL) noises. It has been reported in [1] that due to the construction noises loud enough to cause permanent hearing loss, many construction workers complain that they cannot hear as well as they used to. To protect workers from noise, regulations have been carried out to limit the maximum permissible noise levels for construction works. For example, in Washington state of the USA, the permissible exposure limit allows an 8-hour, full-shift average exposure of 85dBA; and for every 5dBA increase above this level, the allowable exposure time is cut in half [1]. However, as stated in [2, 3], the average SPL generated by most of the construction tasks are over 90dBA, some of them even exceed 95 dB. Workers must use passive hearing protections to avoid hearing damage, which might bring trouble for communicating clearly while working. It would also lead to the increase possibility of accidents for construction works. Traditional passive barrier is usually placed on the noise sound propagation path to prevent it from spreading towards certain direction. But workers who works near around the construction machine will still be affected by the high SPL noise. A better choice is to block the noise at the noise source side to achieve a global noise reduction to form a quieter working condition. If applying the passive noise control method to control the noise at noise source, a fully covered enclosure is required to stop the noise from spreading to any direction, which also heats up the temperature of the engine inside and shorten the life of the machine itself.

In order to maintain the heat dissemination path, the idea of applying active noise control (ANC) technology to build virtual sound barrier (VSB), which could be used to cover a noisy machine [4]. The secondary speakers of the ANC system could be installed either on the opening plane of the rectangular cavity [5, 6, 7] or on the boundary of the opening [8]. For the planar control strategy, it is able to arrange the secondary sources array, and minimize the distance between the adjacent secondary sources, which could enlarge the controllable frequency range. However, the opening will be blocked by the speakers, especially those large volume low-frequency speakers. For the boundary control strategy, the opening is not blocked by the speakers, which maintains the heat dissemination path. The shorter side of the opening plane decides the maximum controllable range of the ANC system [7], which need to be carefully designed based on the target control noise frequency range. 

Comparing with passive solutions, ANC methods could achieve considerably better noise reduction performance at low frequency (below 1000Hz). After analyzing the raw recorded noise data of the often used engine and genset on construction sites, we found that the noise signal energy mostly exists below 500Hz, which indicates ANC is quite promising to control these machinery noises. Aiming to control the real construction machinery noise, an initial stage ANC system applying boundary control strategy is built. The on-site measured sound data indicates that the SPL levels of the machinery noises are above 90dBA. Two kinds of customized low-frequency speakers are designed to generate powerful enough low-frequency anti-noise signal. A high performance Genelec speaker is adopted as the primary source to play back different types of noises.

The rest of this paper is arranged as follows. The basic concept of the ANC system applying boundary control is elaborated in section II.  The details of the designed ANC system targeting on controlling high SPL low-frequency machinery noise are presented in section III. The experimental results are shown in section IV. Conclusions are drawn in section V.

\section{Active noise control system applying boundary control}
The basic concept of ANC technique is generating the anti-signal $\mathbf{Y}(n)$ based on the reference signal $\mathbf{X}(n)$ and error signal $\mathbf{e}(n)$ received by reference and error microphones to attenuate the desired signal $\mathbf{d}(n)$ at monitoring points or areas [9]. The filtered-X least mean square error (FxLMS) algorithm in time domain is extensively used for ANC application for its simplicity and efficiency. The basic block diagram of multi-channel FxLMS algorithm is plotted in Fig. 1. The residual error signals sensed at error microphones could be expressed as 
\begin{equation}
\mathbf{e}(n)=\mathbf{d}(n)-\mathbf{s}(n)*[\mathbf{W}^{T}(n)\mathbf{X}(n)].
\end{equation}
where $\mathbf{s}(n)$ is the impulse response of secondary path $\mathbf{S}(z)$, and $\mathbf{W}(n)$ is the control weight which tries to simultaneously model primary path $\mathbf{P}(z)$ and secondary path $\mathbf{S}(z)$ [9]. The control weight $\mathbf{W}(n)$ is updated as 
\begin{equation}
\mathbf{W}(n+1)=\mathbf{W}(n)+\mu\mathbf{X}^{\prime}(n)\mathbf{e}(n).  
\end{equation}
where $\mu$ is the step size, and $\mathbf{X}^{\prime}=\mathbf{s}(n)\ast\mathbf{X}(n)$. And $\mathbf{s}(n)$ is the impulse response of the secondary path $\mathbf{S}(z)$.

\begin{figure}
	\centering
	\includegraphics[width=8.5cm]{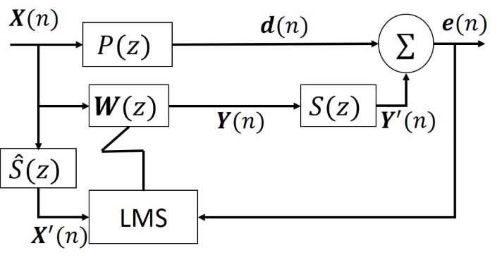}
	\caption{The block diagram of ANC system using FxLMS algorithm.}
	\label{Fig1}
\end{figure}

An VSB system is an array of acoustic sources and sensors forming an acoustic barrier, which blocks the direct propagation of sound without much blocking air, light, and access [4] to create a globally quieter environment outside by applying ANC techniques. The schematic diagram of an VSB system with secondary sources at the boundary of the opening of a rigid cavity is presented in Fig.2. 

\begin{figure}
	\centering
	\includegraphics[width=8.5cm]{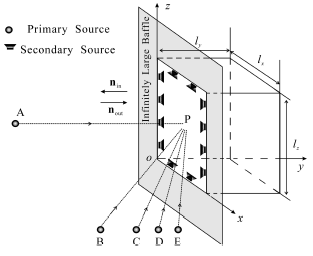}
	\caption{A virtual sound barrier system with the secondary sources at the boundary of the opening of a rigid cavity.}
	\label{Fig2}
\end{figure}

The open cavity is $l_x$ long, $l_y$ wide, and $l_z$ high. When all walls inside the cavity are rigid, the sound pressure inside could be represented as
\begin{equation}
\begin{aligned}
  p_{in} (x,y,z) & =\Sigma_n\Sigma_m[p_{n,m}^{+}e^{-jk_{y,n,m}y}\\
                 & +{p_{n,m}^{-}e^{-jk_{y,n,m}y}}]\dot \phi_{n,m} (x,z)
 \end{aligned}
\end{equation}
where $p_{n,m}^{+}$ and $p_{n,m}^{-}$ correspond to the $(n,m)$th mode in the positive and negative $y$ directions, $k_{y,n,m}=\sqrt{k^2 - (n\pi/l_x)^2 - (m\pi/l_z)^2}$ is the wave number in the $y$ directions, and $\phi_{n,m}(x,z)=cos(n\pi x/l_x)cos(m\pi z/l_z)$ is the eigenfunction of the $(n,m)$th mode. And the acoustic pressure outside the cavity is the summation of the incident pressure $p_i$, the reflection pressure $p_r$, and the scattering pressure $p_{sca}$
\begin{equation}
  p_{out} (x,y,z)  =p_{i} (x,y,z)+p_{r} (x,y,z)+p_{sca} (x,y,z)
\end{equation}
And the sound pressure generated by the secondary sources at boundary is
\begin{equation}
\begin{aligned}
  p_{s} (x,y,z) & =\Sigma_n\Sigma_m[p_{s,n,m}^{+}e^{-jk_{y,n,m}y}\\
                & +{p_{s,n,m}^{-}e^{-jk_{y,n,m}y}}]\phi_{n,m}(x,z)\\
                & + {{\iiint}_{V}}{j{{\rho}_0}{\omega}q({\mathbf{r}}_s)G_A{dV}}.
 \end{aligned}
\end{equation}
where $p_{n,m}^{+}$ and $p_{n,m}^{-}$ correspond to the $(n,m)$th mode in the positive and negative $y$ directions respectively, and $\mathbf{r}_s$ is the location of the secondary source, $G_A$ is the Green's function. The cost function for optimizing the secondary source strength is 
\begin{equation}
  J =\frac{1}{2{\rho}{c_0}^2}|{p_t}|^2,
\end{equation}
where $p_t$ is the total sound pressure at evaluation with control. The optimized strengths of the secondary sources are 
\begin{equation}
  \mathbf{q}_S=(\mathbf{S}^H(z)\mathbf{S}(z))^{-1}\mathbf{S}(z)\mathbf{P}(z)q_p,
\end{equation}
where $q_p$ is the source strength of the primary source. The performance of the boundary ANC control system is evaluated with ratio of the summation of the squared sound pressure in the target area without $p_p(\mathbf{r}_{V,i})$ and with $p_{t,o}(\mathbf{r}_{V,i})$ control.
\begin{equation}
  NR=10log_{10}\frac{\sum_{i=1}^{N_V}|p_p(\mathbf{r}_{V,i})|^2}{\sum_{i=1}^{N_V}|p_{t,o}(\mathbf{r}_{V,i})|^2},
\end{equation}
where $N_V$ is the number of evaluation points.

\begin{figure*}[ht]
	\centering
	\includegraphics[max size={\textwidth}{\textheight}]{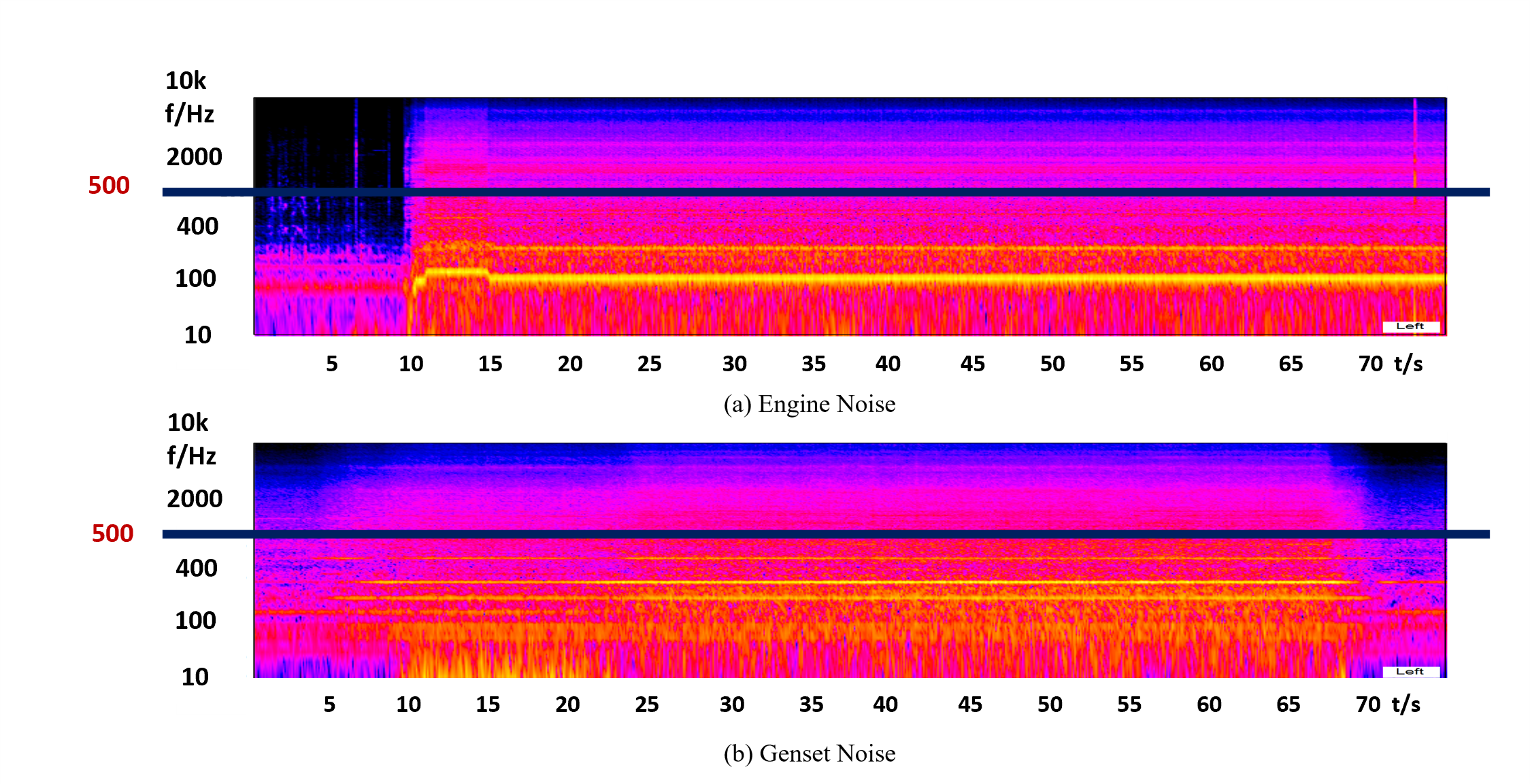}
	\caption{Spectrogram of construction machinery noises.}
	\label{Fig3}
\end{figure*}

\section{Experimental setup}

Engine and genset machines are quite commonly used on construction sites. The SPL could reach over 95 dB, which may cause irreversible hearing loss for construction worker working around. These machinery noises at construction site using high-quality recording microphones are recorded for around one minute. The SPL of these two machinery noises are both over 95dB. The corresponding spectrogram are plotted in Fig.3. The signal energy of both machinery noises concentrates below 500Hz, which is within the main controllable frequency of ANC techniques.

In order to control the low-frequency machinery noises in an efficient way, a four-channel ANC system, which forms a boundary control VSB, is designed and built. The schematic diagram of this ANC system is plotted from overall view (a), top view (b), and side view (c) in Fig.4. The real system setup picture is presented in Fig.5. The enclosure with opening is 64cm long, 50cm wide, and 92cm high. A high performance Genelec speaker is placed inside the enclosure as the primary source to play back the soundtracks for different kinds of noises.

\begin{figure}
	\centering
	\includegraphics[max size={\textwidth}{\textheight}]{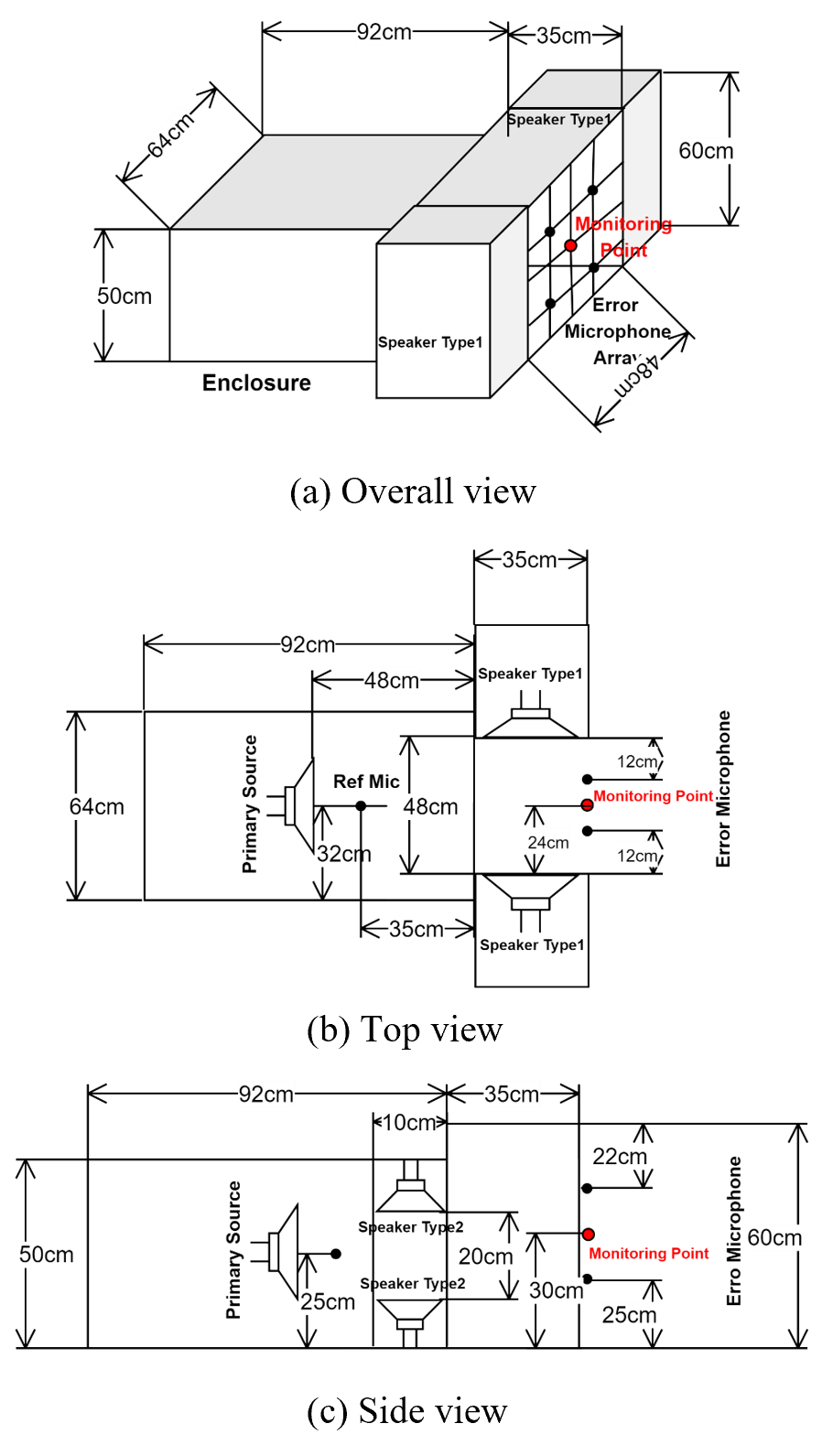}
	\caption{The schematic diagram of the boundary control VSB system.}
	\label{Fig4}
\end{figure}

\begin{figure*}
	\centering
	\includegraphics[max size={\textwidth}{\textheight}]{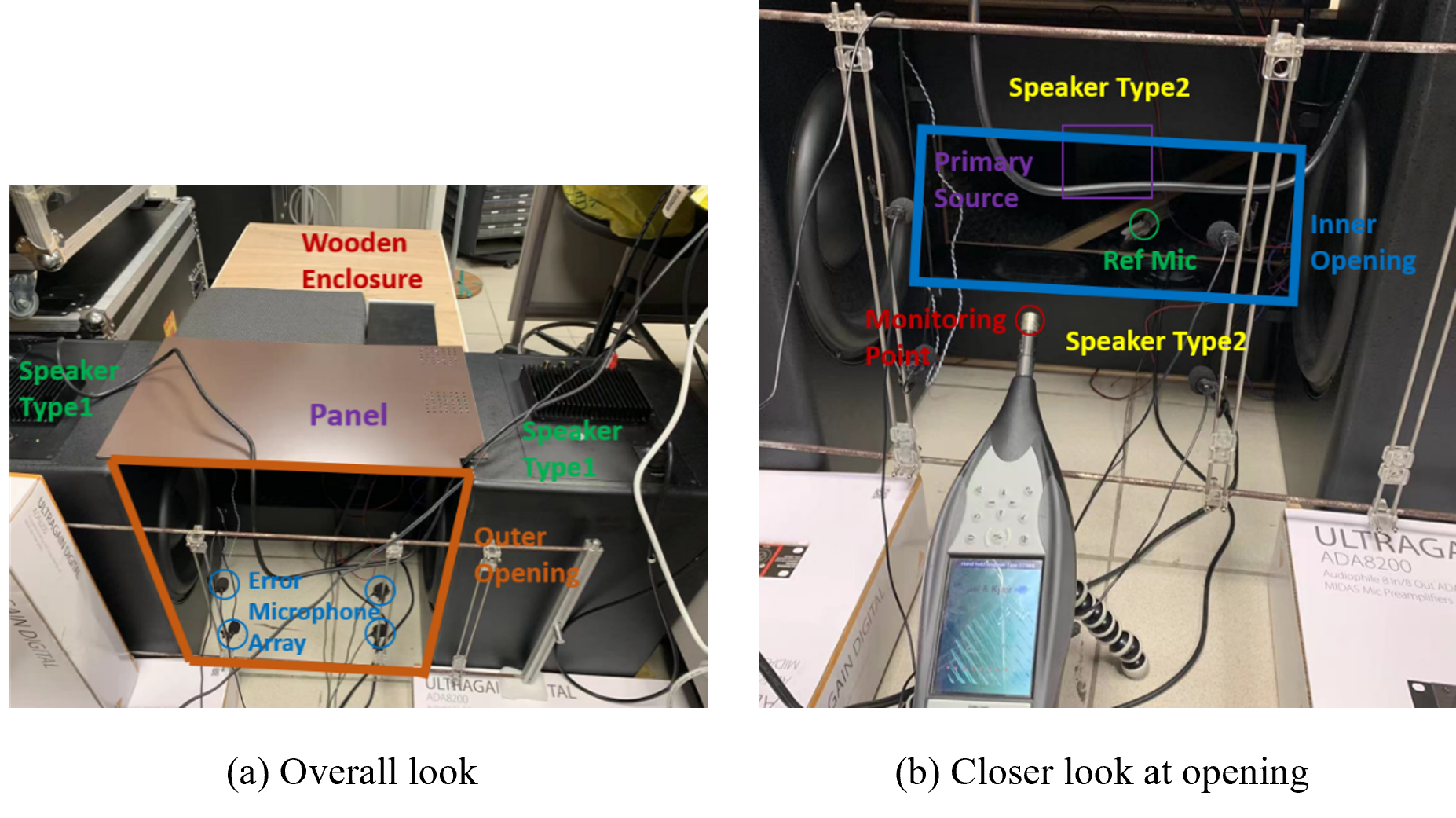}
	\caption{The picture of the real boundary control VSB system setup.}
	\label{Fig5}
\end{figure*}

Two types of customized low-frequency speakers are designed and manufactured, desiring to control the low-frequency machinery noise, which are shown in Fig.6. The Speaker A is targeting on generating high-power output low-frequency sound below 500Hz. The Speaker B assist Speaker A to control the high-frequency components higher than 500Hz. Panels and two Speaker A are adopted to form a duct at the opening of enclosure. Two Speaker B are mounted against the opening of the enclosure, which can be seen from Fig.4(c) and Fig.5(b). The size of the inner opening formed by four secondary source speakers is $20cm \times 48cm$. And the size of the outer opening formed by two speaker type 1 and two panel $48cm \times 60cm$. The higher limitation of the controllable frequency range of this boundary control VSB system is determined by the shorter side $l_s$ of the opening. The wavelength of the highest controllable frequency ${\lambda}_{h}=\frac{c}{f_{h}}$ is less than the $2\times{{l}_s}$ [10]. Thus, for this boundary control VSB system, the highest controllable frequency is $f_{h}=\frac{c}{2l_{S}}=875Hz$. However, the frequency responses of secondary sources also determine the controllable range of the ANC system. The frequency responses of these two speakers are presented in Fig.7. The impedance characteristics of speaker A is provided by the speaker manufacturer. That of speaker B is not provided. We only concern the frequency response of the speakers here. As we can see, the mainly concerned frequency range of speaker type 1 is from 30Hz to 500Hz with around 100dB output, and that of speaker type 2 is from 80Hz to 1000Hz with around 80dB output. Thus, for high SPL primary noise, we are mainly focusing on the frequency range 30$\sim$500Hz.

\begin{figure}
	\centering
	\includegraphics[max size={\textwidth}{\textheight}]{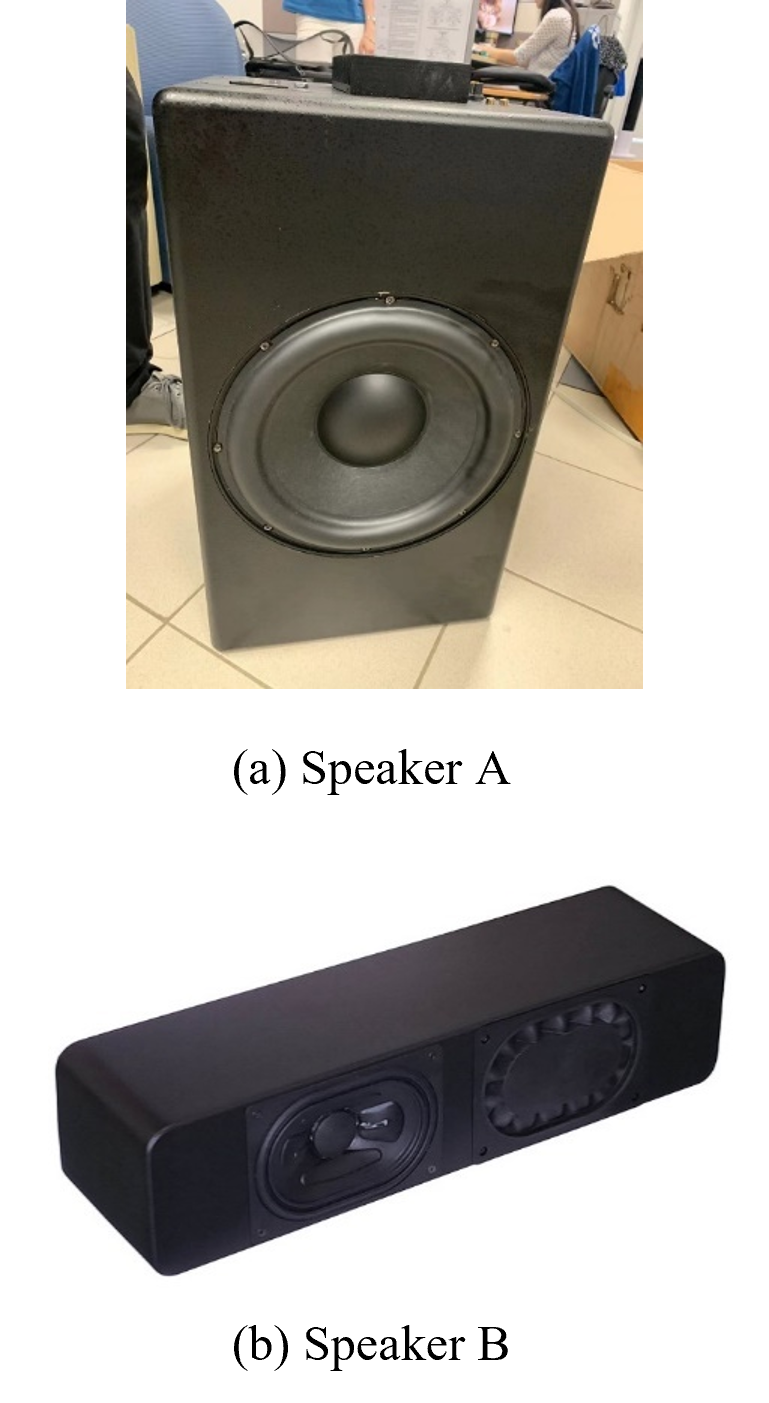}
	\caption{The outlook of the two customized low-frequency speakers.}
	\label{Fig6}
\end{figure}

\begin{figure}
	\centering
	\includegraphics[max size={\textwidth}{\textheight}]{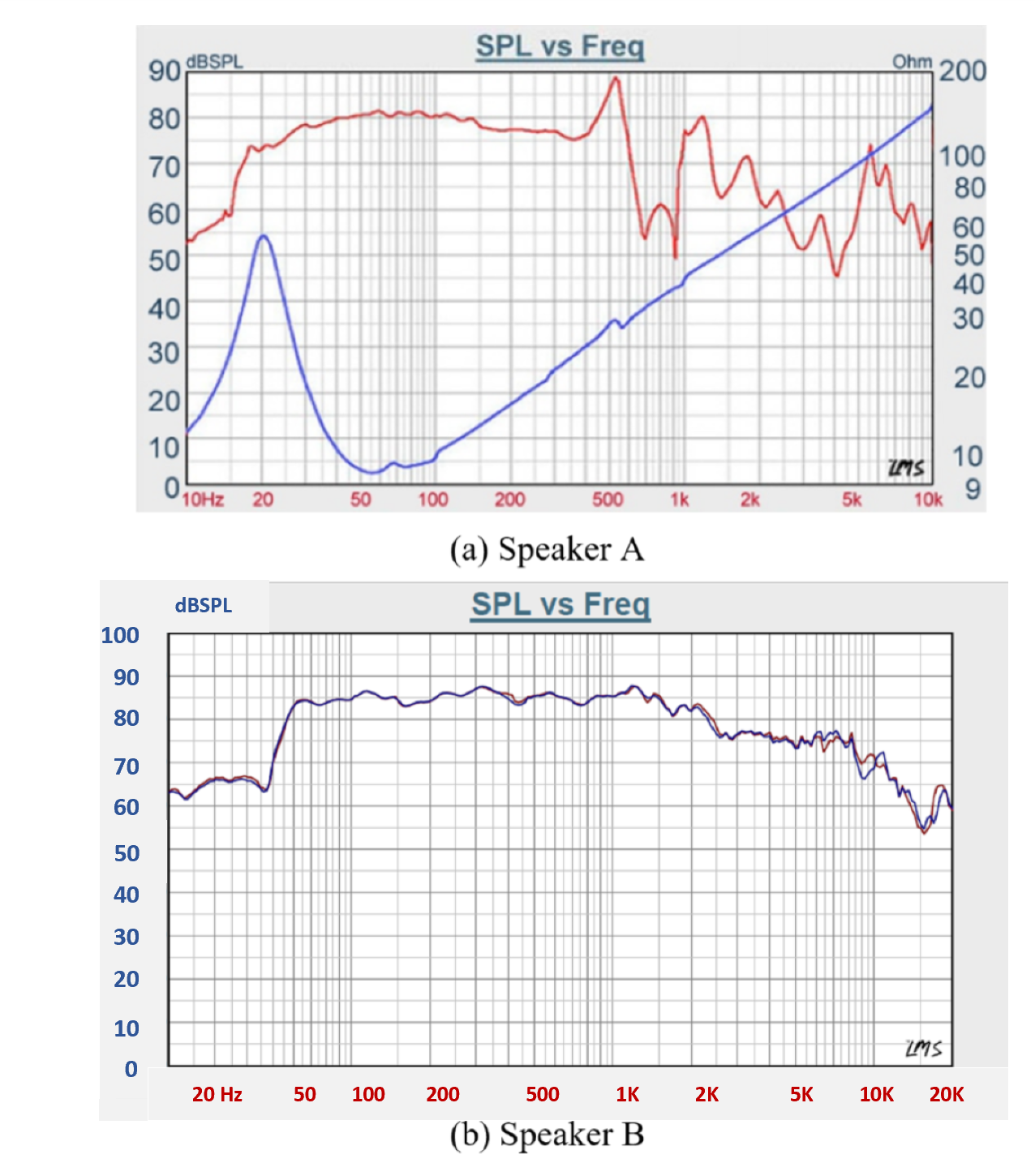}
	\caption{The frequency response of the two customized low-frequency speakers.}
	\label{Fig7}
\end{figure}

\section{Experimental Results}
The TMS320C6713-DSK board is adopted as the main controller board of the ANC system, where the centralized multi-channel feedforward FxLMS algorithm is implemented. The control filter and secondary path are both 128 taps. The sampling rate is set as 16kHz, which is the rated sampling rate of TMS320C6713-DSK board. The step size of FxLMS algorithm in time domain is set as $10^{-9}$. The SPL of background noise is 40dB.

\subsection{Tonal Noise}
In order to test out the active control ability of the system, tonal noises from 50Hz to 650Hz with 50Hz interval are tested. The noise reduction performances for the tonal noises are summarized in Fig.8. As we can see, for tonal noises from 150Hz to 600Hz, the reduction is around 20$\sim$25dB. Around 10dB reduction is gained at 50Hz. However, the noise control performance at 100Hz is quite limited, where the mode of this open enclosure lies. For tonal noises of which the frequency are over 600Hz, no noise reduction could be gained at all. The tonal noise control results meets our expectations stated in Section 3.

\begin{figure}
	\centering
	\includegraphics[max size={\textwidth}{\textheight}]{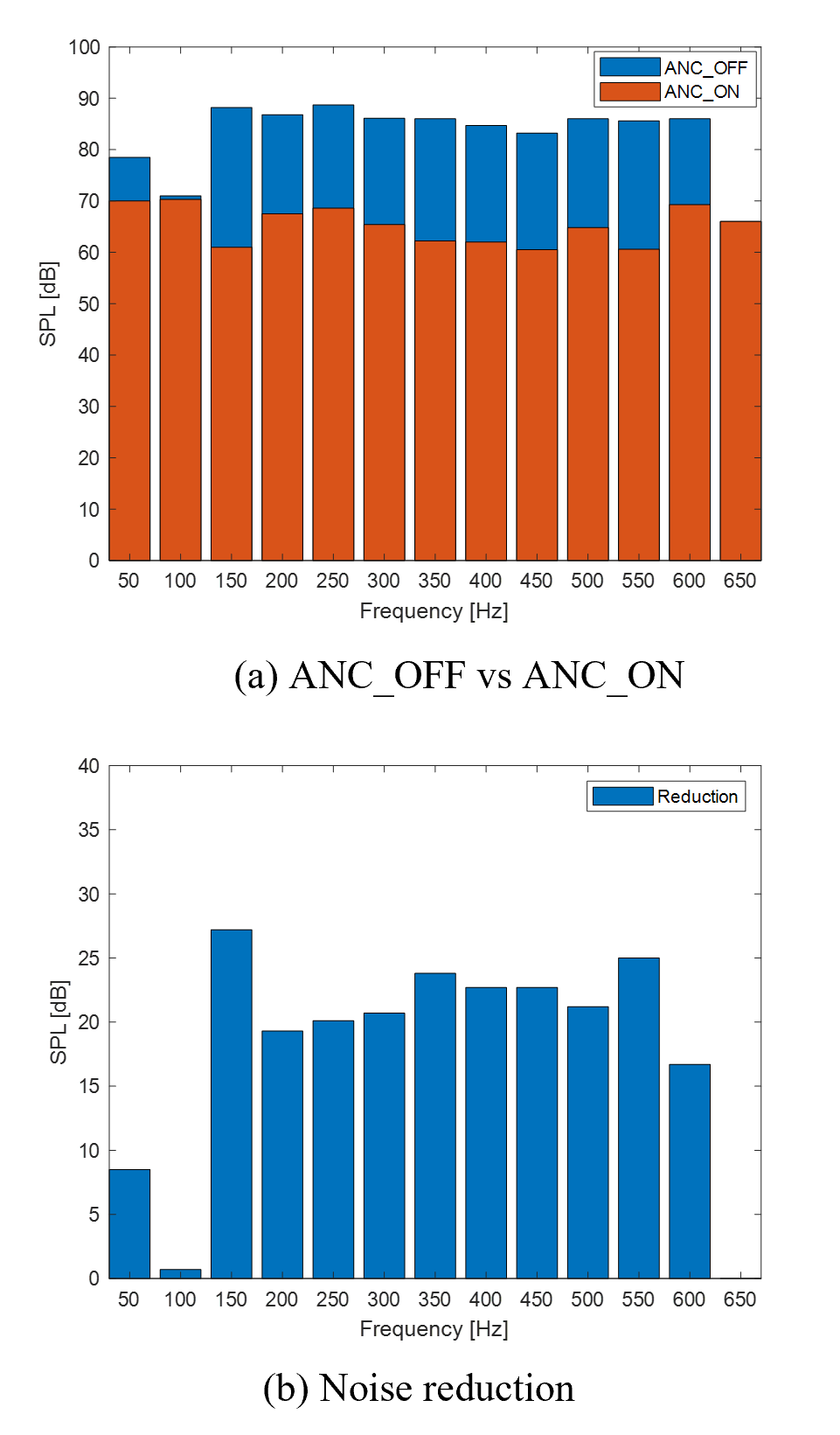}
	\caption{The noise reduction performance of the ANC system for tonal noise.}
	\label{Fig8}
\end{figure}

\subsection{Broadband Noise}
As show in Fig.3, the practical machinery noises are broadband noise with multiple highlighted bands below 500Hz. The tonal noise control results presented in Fig.8 already prove that the frequency control range of this ANC system covers the main highlighted frequency bands. Thus, it is quite promising to have a good results on controlling machinery noises. In order to investigate how does this ANC system perform on broadband noise control, White Gaussian Noise (WGN) with two different bands, breaker and air compressor noises downloaded online, and raw recorded engine and genset noises are adopted as primary noises. The high-performance Genelec speaker inside the cavity plays back one of these broadband noise each time. The corresponding control results are summarized in Table 1. 
\begin{table}
\begin{center}
\begin{threeparttable}
\caption{Noise reduction performance of the ANC system for different broadband noises.}
\begin{tabular}{|c|c|c|c|}
\cline{1-4}
Type of Noise & ANC\_OFF & ANC\_ON & Noise Reduction \\
\cline{1-4}
150$\sim$600Hz WGN & 87dB & 80dB & 7dB \\
\cline{1-4}
200$\sim$300Hz WGN  & 91dB & 76dB & 15dB \\
\cline{1-4}
Compressor Noise & 89.7dB & 81dB & 8.7dB\\
\cline{1-4}
Breaker Noise & 89.6dB & 79.1dB & 10.5dB\\
\cline{1-4}
Engine Noise & 96.5dB & 87.2dB & 9.3dB \\
\cline{1-4}
Genset Noise & 97.6dB & 93.2dB & 4.4dB\\
\hline
\end{tabular}
\end{threeparttable}
\end{center}
\end{table}

As we can see, for those broadband noises, of which the frequency components of the main signal energy concentrates on one narrow band at low-frequency part, such as breaker, engine and air compressor noises, around 10 dB reduction could be gained. The engine and genset noises control results are compared in detail. The high-precision GRASS microphone is used to record the signal at monitoring point when ANC is OFF or ON. The 1/3 octave bands of the recorded signals are presented in Figure 9 and 10. The engine noise has an evident peak around 50Hz band, of which the SPL is 15dB higher than the rest signal. This peak is clearly suppressed by the ANC system. Thus, the overall noise reduction for engine noise is 9.3dB. However, the overall noise reduction for genset noise is only 4.4 dB, which is not good enough. Only the signal energy peak around 200Hz got obvious attenuated, the signal energy peak around 50Hz is barely reduced. This is resulted by the limited computational resources and non-ideal analog-digital conversion speed of the main processing board. 
\begin{figure*}[ht]
	\centering
	\includegraphics[max size={\textwidth}{\textheight}]{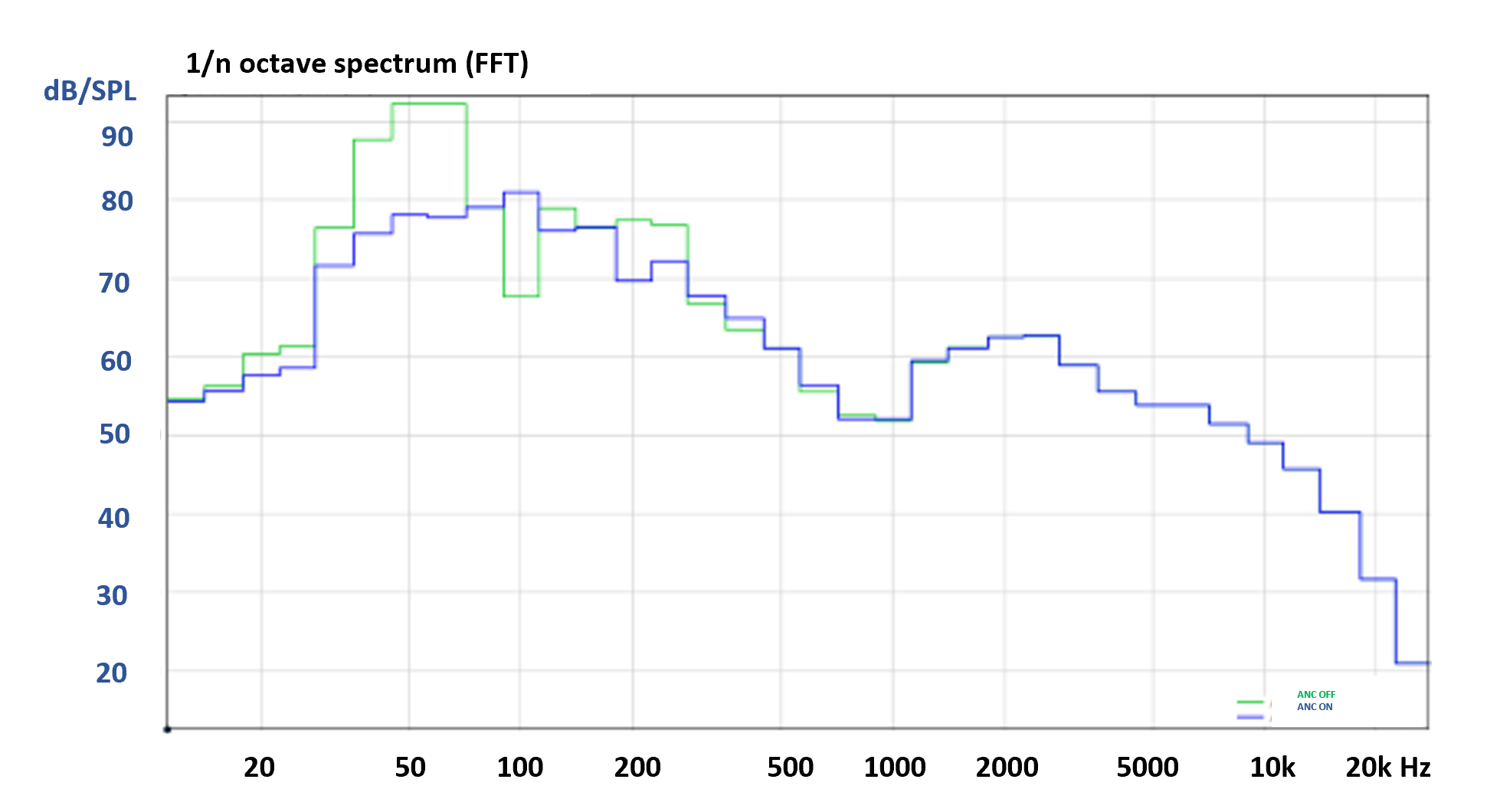}
	\caption{The noise reduction performance of the ANC system for Engine noise.}
	\label{Fig9}
\end{figure*}

\begin{figure*}[ht]
	\centering
	\includegraphics[max size={\textwidth}{\textheight}]{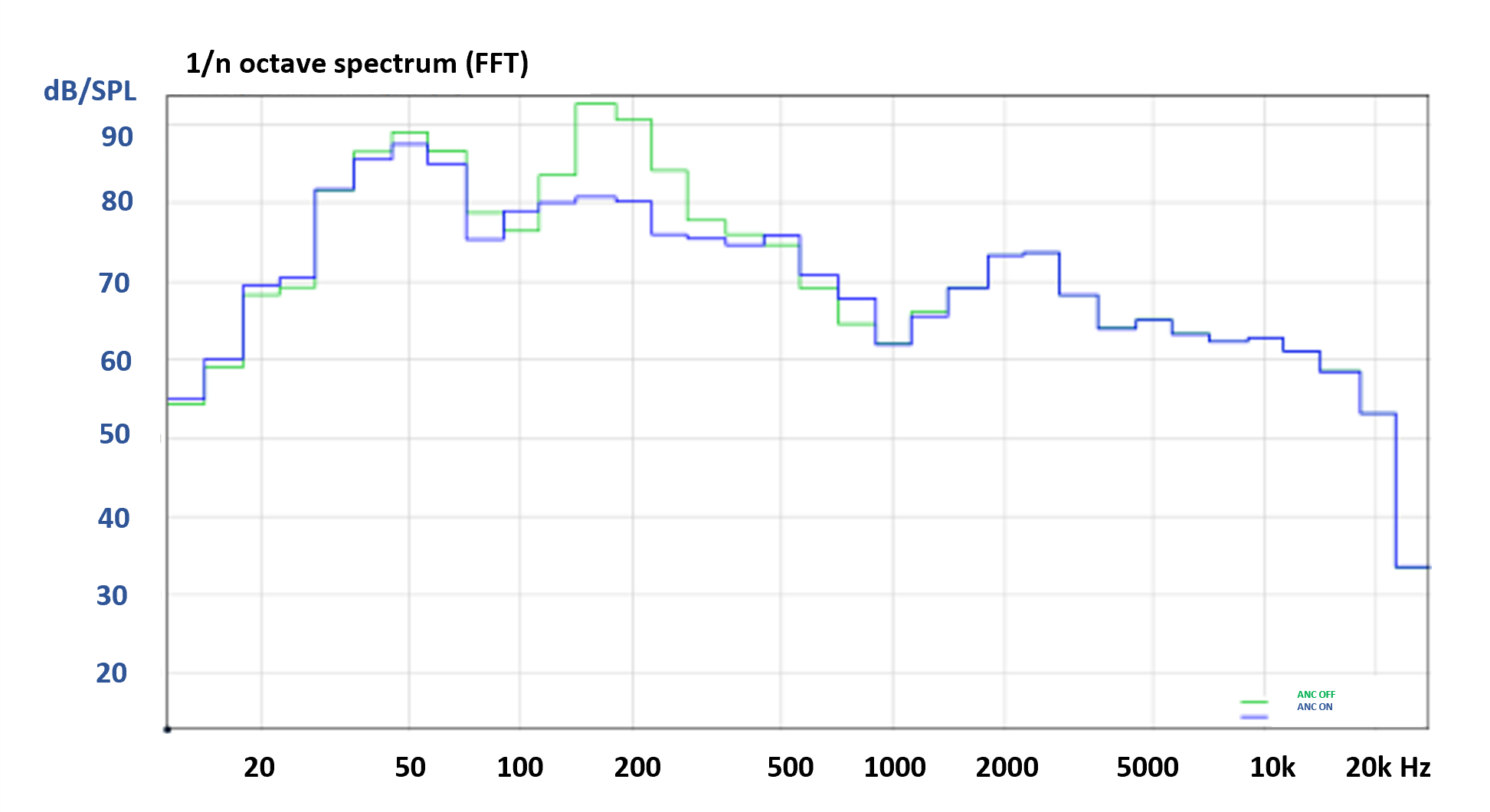}
	\caption{The noise reduction performance of the ANC system for Genset noise.}
	\label{Fig10}
\end{figure*}

\section{Conclusions}
A four-channel ANC system applying boundary control is designed for low-frequency practical machinery noise control. Two types of customized low-frequency high output power speakers are designed and manufactured for this application scenario.The TMS320C6713-DSK board is adopted as the main processing board of the ANC system to implement multi-channel FxLMS algorithm in time domain. Tonal noises from 50Hz to 600Hz, except for 100Hz, could be well controlled. Practical broadband noises of which the frequency components of the main signal energy concentrates on one narrow band at low-frequency part, such as breaker, engine and air compressor noises, around 10 dB reduction could be gained. Thus, half of the perceived loudness has been reduced. The permissible working hour for the construction workers could be lengthen four times longer, which could speed up noisy construction works.

\section*{Acknowledgment}
This material is based on research supported by the Singapore Ministry of National Development and National Research Foundation under the Cities of Tomorrow R\&D Programme: COT-V4-2019-1. Any opinions, findings, and conclusions or recommendations expressed in this material are those of the author(s) and do not reflect the views of the Singapore Ministry of National Development and National Research Foundation, Prime Minister’s Office, Singapore.

\end{document}